\documentclass[aps,twocolumn,showpacs,showkeys,nofootinbib,superscriptaddress]{revtex4-1}
\usepackage{graphicx}     
\usepackage{amsmath}
\usepackage{amssymb, amsfonts}
\usepackage{array}
\newcolumntype{x}[1]{>{\centering\let\newline\\\arraybackslash\hspace{0pt}}p{#1}}
\usepackage{braket}
\usepackage{cases}
\usepackage[utf8]{inputenc}
\usepackage[overload]{empheq}
\usepackage{relsize}  
\usepackage{multirow}
\usepackage{dcolumn}  
\usepackage{hyperref}
\usepackage{xcolor}
\usepackage{relsize}%
\hypersetup{
	colorlinks   = true, 
	urlcolor     = blue, 
	linkcolor    = blue, 
	citecolor    = blue 
}
\newcommand{\RN}[1]{%
	\textup{\uppercase\expandafter{\romannumeral#1}}%
}
\newcommand{\angstrom}{\text{\normalfont\AA}}
\newcolumntype{d}[1]{D{.}{.}{#1}}
\newcolumntype{K}[1]{>{\centering\arraybackslash}p{#1}}

\usepackage{floatflt}     
\usepackage{wrapfig}
\usepackage{color}
\usepackage{pstricks}

\makeatletter
\newcommand*{\rom}[1]{\expandafter\@slowromancap\romannumeral #1@}
\usepackage{color}
\definecolor{DarkRed}{rgb}{0.35,0.01,0.01}
\definecolor{Linen}{rgb}{0.98,0.98,0.94}
\definecolor{Blue}{rgb}{0.,0.,1.0}
\definecolor{DarkBlue}{rgb}{0.099,0.099,0.44}
\definecolor{DarkGreen}{rgb}{0.0,0.4,0.0}
\definecolor{Turquoise}{rgb}{0.0,0.9,0.7}
\begin{document}

\title{Lateral transition metal dichalcogenide heterostructures for high efficiency thermoelectric devices}

\author{Sathwik Bharadwaj}
\affiliation{Department  of   Physics,  Worcester
	Polytechnic Institute, Worcester, Massachusetts 01609, USA.}
\affiliation{Center for Computational NanoScience,  Worcester
	Polytechnic Institute, Worcester, Massachusetts 01609, USA.}
\author{Ashwin Ramasubramaniam}
\email{ashwin@engin.umass.edu}
\affiliation{Department of Mechanical and Industrial Engineering, University of Massachusetts, Amherst, Massachusetts 01003, USA.}
\author{L. R. Ram-Mohan}
\email{lrram@wpi.edu}
\affiliation{Department  of   Physics,  Worcester
	Polytechnic Institute, Worcester, Massachusetts 01609, USA.}
\affiliation{Center for Computational NanoScience,  Worcester
	Polytechnic Institute, Worcester, Massachusetts 01609, USA.}

\begin{abstract}
Increasing demands for renewable sources of energy has been a major driving force for developing efficient thermoelectric materials. Two-dimensional (2D) transition-metal dichalcogenides (TMDC) have emerged as promising candidates for thermoelectric applications due to their large effective mass and low thermal conductivity. In this article, we study the thermoelectric performance of lateral TMDC heterostructures within a multiscale quantum transport framework. Both $n$-type and $p$-type lateral heterostructures are considered for all possible combinations of semiconducting TMDCs: MoS$_2$, MoSe$_2$, WS$_2$, and WSe$_2$. The band alignment between these materials is found to play a crucial in enhancing the thermoelectric figure-of-merit ($ZT$) and power factor far beyond those of pristine TMDCs. In particular, we show that the room-temperature $ZT$ value of $n$-type WS$_2$ with WSe$_2$ triangular inclusions, is five times larger than the pristine WS$_2$ monolayer. $p$-type MoSe$_2$ with WSe$_2$ inclusions is also shown to have a room-temperature $ZT$ value about two times larger than the pristine MoSe$_2$ monolayer. The peak power factor values calculated here, are the highest reported amongst gapped 2D monolayers at room temperature. Hence, 2D lateral TMDC heterostructures open new avenues to develop ultra-efficient, planar thermoelectric devices. 
\end{abstract}
\maketitle
\begin{center}
\today
\end{center}
Thermoelectric devices can play a pivotal role in fulfilling future demands for clean energy \cite{Dresselhaus1, Dresselhaus2, vineis}. A good
thermoelectric material must have a high thermoelectric figure-of-merit $ZT$, defined as
\begin{equation}\label{eq:zt}
ZT = \frac{\sigma S^2 T}{\kappa_e +\kappa_{ph}},
\end{equation}
where $T$ is the absolute temperature, $\sigma$ is the electrical conductance, $S$ is the Seebeck coefficient, $\kappa_e$ is the electronic thermal conductivity, and $\kappa_{ph}$ is the lattice phonon thermal conductivity. In bulk materials, the $ZT$ value is limited by $\sigma$ and $S$ varying in inverse proportion, and $\kappa_e$ and $\sigma$ varying in direct proportion (Wiedemann-Franz law)  \cite{Wiedemann}. Hence, for a long period of time thermoelectricity was
believed to be an inefficient source of energy for practical application \cite{rowe}. However, works by
Hicks and Dresselhaus \cite{hicks1,hicks2,hicks3} illustrated that in nanostructures, one could achieve a substantial increase in the value of $ZT$ by reducing the dimensionality of the system. The density of electronic states per unit volume increases in lower dimensions, thereby resulting in an enhancement in $ZT$ \cite{hung}. Since then the field of thermoelectricity has focused on: a)
increasing $S$ and $\sigma$ independently through quantum confinement effects, and b) decreasing
$\kappa_{ph}$ by systematically controlling phonon contributions \cite{majumdar}.
Additionally, other techniques such as band-gap
engineering \cite{pei}, carrier-pocket engineering \cite{Koga}, energy filtering \cite{Bahk}, and
semimetal–semiconductor transition \cite{LinRobin} have been developed to engineer the thermoelectric properties of nanostructures.

Traditionally, semiconductor superlattices and heterostructures have been used to construct efficient thermoelectric devices. However, in such structures, it is experimentally difficult to achieve the efficiency predicted by theory since a large number of parameters have to be optimized \cite{Zebarjadi}. In this regard, two-dimensional
(2D) materials such as graphene and transition-metal dichalcogenides (TMDC) have attracted tremendous attention due to their unique physical and chemical properties \cite{Bonaccorso}. The high degree of flexibility of 2D materials to tune the electrical and thermal properties, makes them ideal candidates for thermoelectric applications. The prototypical 2D material, graphene, has exhibited a power factor (PF) value as high as $34.5\,$mWm$^{-1}$K$^{-2}$ at room temperature \cite{Andrei}. However, it has limited thermoelectric applications due to an extremely high thermal conductivity (\mbox{$2000-4000\,$Wm$^{-1}$K$^{-1}$} for freely suspended samples at room temperature \cite{Ruoff1,Ruoff2}). In comparison, monolayer (1L) TMDCs maintain a very low thermal conductance due to significantly lower phonon mean-free paths \cite{cai,Ozbal}. Hence, TMDCs have tremendous potential to realize in-plane thermoelectric and Peltier cooling devices. 

There have been several first-principles studies on calculating the thermoelectric quantities in 1L and layered TMDCs \cite{HuangTMD1, HuangTMD2, lake, Babaei, Ouyang, Hippalgaonkar, Lundstrom}. $p$-type MoS$_2$ 1L and $n$-type WSe$_2$ 1L were observed to have maximum $ZT$ values at room temperature and at higher temperatures, respectively. Also, bilayer MoS$_2$ was observed to have a PF of $8.5\,$mWm$^{-1}$K$^{-2}$, which is the highest amongst materials with a non-zero bandgap \cite{Hippalgaonkar}. Yet the conductance and $ZT$ values observed in TMDCs are much lower than the corresponding quantities in traditional thermoelectric materials such as Bi$_2$Te$_3$, and phonon-glass electron-crystals \cite{Ertekin}. Thus, there are opportunities to boost the thermoelectric performance in TMDCs through the formation of heterostructures, as delineated in this article.

Similar crystal structure and comparable lattice constants of MX$_2$ (M = Mo, W; X = S, Se) monolayers have motivated the construction of lateral TMDC heterostructures. Experimentally, such structures are fabricated through multistep chemical vapor deposition techniques \cite{growth3, growth4, growth5, growth6, growth7}, one-pot synthesis \cite{growth8}, and omnidirectional epitaxy \cite{growth9}. In traditional thermoelectric materials, such as Bi$_2$Te$_3$, quantum confinement through the formation of heterostructures has been demonstrated to enhance the figure-of-merit \cite{yuchi, Venkatasubramanian}. Thus, we may anticipate such an enhancement in lateral 2D TMDC heterostructures as well.

\begin{figure}[th]
	\centering
	\includegraphics[scale = 0.1]{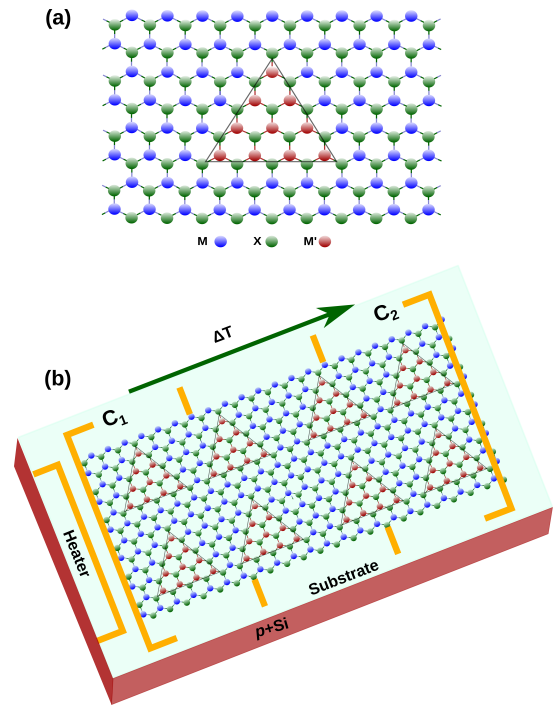}
	\caption{A schematic representation of (a) a triangular inclusion in 2D materials is displayed. Here, M$^{\prime}$X$_2$ material is confined within the MX$_2$ matrix. (b) An in-plane thermoelectric device using lateral TMDC heterostructures is shown. The 2D monolayer is placed on an oxide substrate that can be grown on $p+$silicon material.}
	\label{fig:2dqd}
\end{figure}

In this article, we study the thermoelectric performance of lateral TMDC heterostructures within a multiscale quantum transport framework with inputs from first-principles calculations. We specifically consider triangular inclusions (see Fig.~\ref{fig:2dqd}), since 2D TMDCs are typically grown as triangular flakes. We study both $n$-type and $p$-type lateral heterostructures, for all possible combinations of semiconducting TMDC monolayers: MoS$_2$, MoSe$_2$, WS$_2$, and WSe$_2$. $n$-type WS$_2$ with WSe$_2$ triangular inclusions is found to have $ZT \approx 1$ at room temperature, which is five times larger than the $ZT$ value of pristine $n$-type 1L WS$_2$. The peak room-temperature power factors calculated for lateral TMDC heterostructures here are the highest amongst gapped 2D monolayers reported to date.


\section{Results}\label{results}
\subsection{Nature of scattering in TMDC heterostructures}
\begin{figure*}[t]
	\centering
	\includegraphics[width=5in]{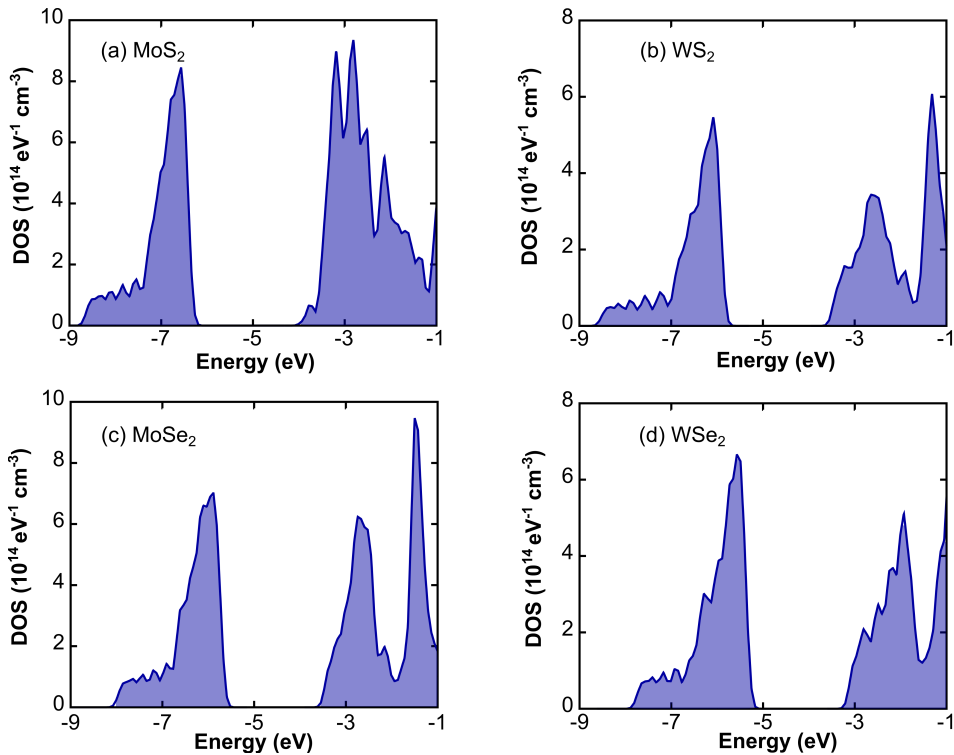}
	\caption{Density of states per unit energy per unit area obtained from DFT calculations is plotted as a function of energy for 1L (a) MoS$_2$, (b) WS$_2$, (c) MoSe$_2$, and (d) WSe$_2$. DFT calculations were performed within local-density approximations (LDA), and the spin-orbit effects are neglected. Hence, all bands are doubly degenerate. The zero of the energy scale is at the vacuum. Bandgaps observed here were corrected to match the GW calculations.}
	\label{fig:dos}
\end{figure*}

We calculated thermoelectric properties using the Boltzmann transport theory under the relaxation time approximations. Within this framework, the kinetic definitions of the conductance, Seebeck coefficient, and the electrical thermal conductivity are given by
\begin{align}\label{eq:thermoelectric}
\sigma &= e^2 \mathcal{I}_0,\nonumber\\
S &= \frac{1}{eT} \frac{\mathcal{I}_1}{\mathcal{I}_0}, \\
\kappa_e &= \frac{1}{T}\left[\mathcal{I}_2-\frac{\mathcal{I}^2_1}{\mathcal{I}_0}\right],\nonumber
\end{align}
with
\begin{equation}
\mathcal{I}_n = \int dE\,v^2\,\tau(E)\,g(E)\,\left(E-\mu_F\right)^n\,\left(-\frac{\partial f_0}{\partial E}\right),
\end{equation}
where $e$ is the elementary charge, $g(E)$ is the density of states, \mbox{$v=\left|\nabla_k E_n(k)\right|/\hbar$} is the carrier velocity, $\displaystyle f_0(E) = 1/(1+e^{(E-\mu_F)/k_B T})$ is the Fermi-Dirac distribution function, $\mu_F$ is the Fermi level, and $\tau(E)$ is the total scattering time. The density of states $g(E)$ is extracted from the electronic band structure obtained using the density functional theory (DFT) calculations within the local-density approximations (LDA). Figure~\ref{fig:dos} displays the density of states as a function of energy for 1L MoS$_2$, WS$_2$, MoSe$_2$ and WSe$_2$. Here, $g(E)$ is normalized by the unit-cell area and the corresponding layer thickness. The PF and $ZT$ values are sensitive to the small variations of $g(E)$ near the band edges. 

To determine $\tau(E)$, we need to consider both the intrinsic and extrinsic scattering rates. According to the Matthiessen’s law
\begin{equation}\label{eq:Matthiessen}
{\tau(E)}^{-1} = {\tau_e(E)}^{-1} + {\tau_{ph}(E)}^{-1},
\end{equation}
where $\tau_e$ is the extrinsic carrier scattering time arising from the material inclusions, and $\tau_{ph}$ is the total intrinsic scattering time arising from all the acoustic and optical phonon mode contributions. The intrinsic scattering rate $\tau_{ph}$ is assumed to remain unaltered from the pristine 1L, a commonly used assumption while studying nano-structured thermoelectric materials \cite{grossman}.   
\begin{figure}[th]
	\centering
	\includegraphics[width=2.5in]{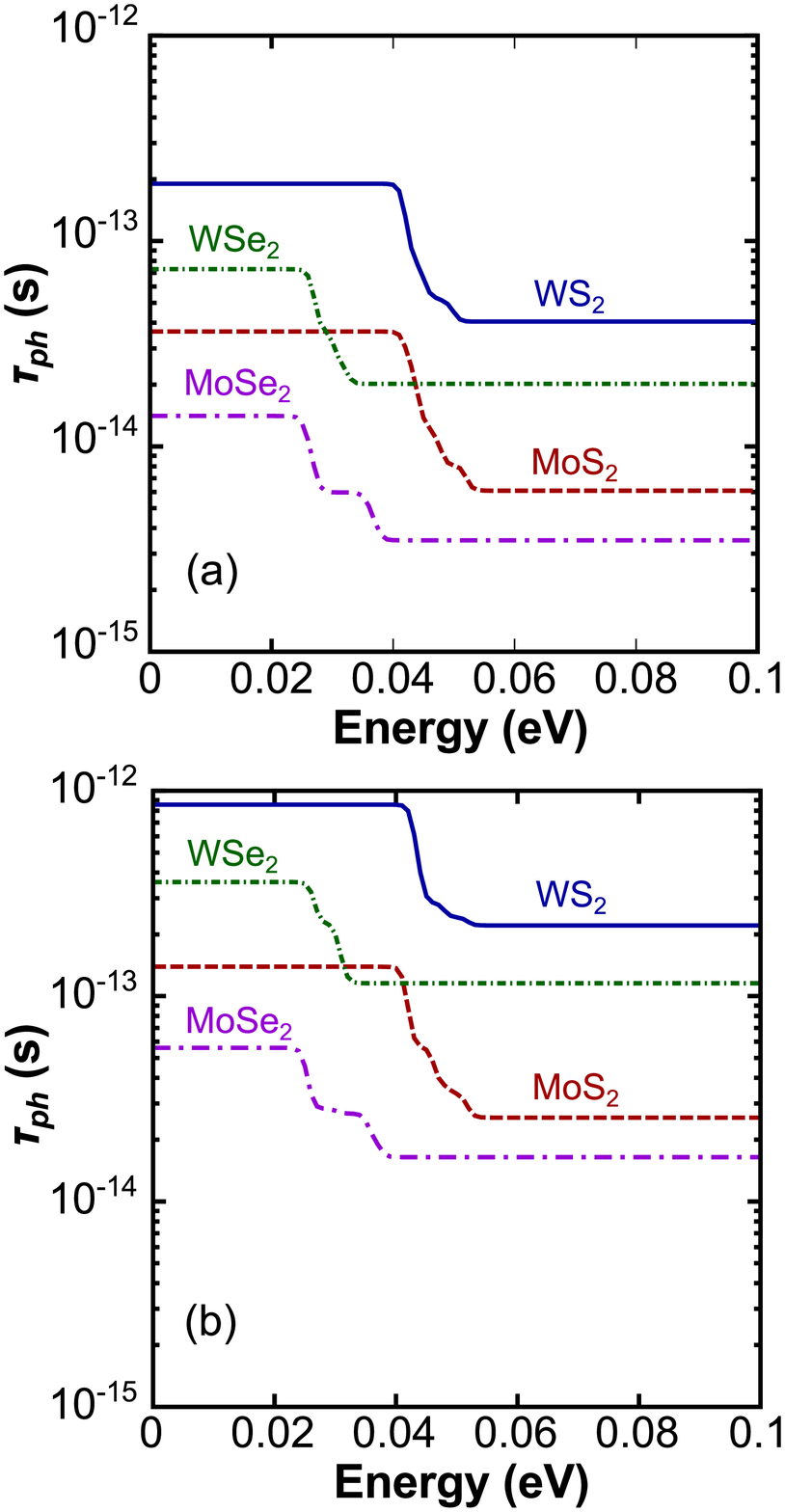}
	\caption{The total phonon scattering time versus energy for (a) $K$-valley electrons, and (b) $K$-valley holes are plotted for TMDCs at temperature $T= 300\,$K. Scattering times are calculated using the deformation potentials listed in Ref.~\cite{Kaasbjerg, kim_eph, kim_eph_mos2}.}
	\label{fig:phononrate}
\end{figure}

Figure~\ref{fig:phononrate} shows the total phonon scattering time versus energy at room temperature for pristine $n$-type and $p$-type 1L TMDCs. We have included the acoustic and optical phonon modes corresponding to the transitions, \mbox{$K\,\rightarrow\,\left\{K, K', Q, Q'\right\}$}, via corresponding zeroth and first-order deformation potentials (see Methods for details). The optical phonon modes emerge as steps in the scattering rate.  In the family of 1L TMDCs, MoSe$_2$ (WS$_2$) has the strongest (weakest) interaction with phonons. In general, WX$_2$ has a greater phonon-limited electrical conductivity than MoX$_2$. These observations and scattering times are consistent with other first-principles studies reported in the literature \cite{Kaasbjerg, kim_eph}. 
\begin{figure}
	\centering
	\includegraphics[width=3.4in]{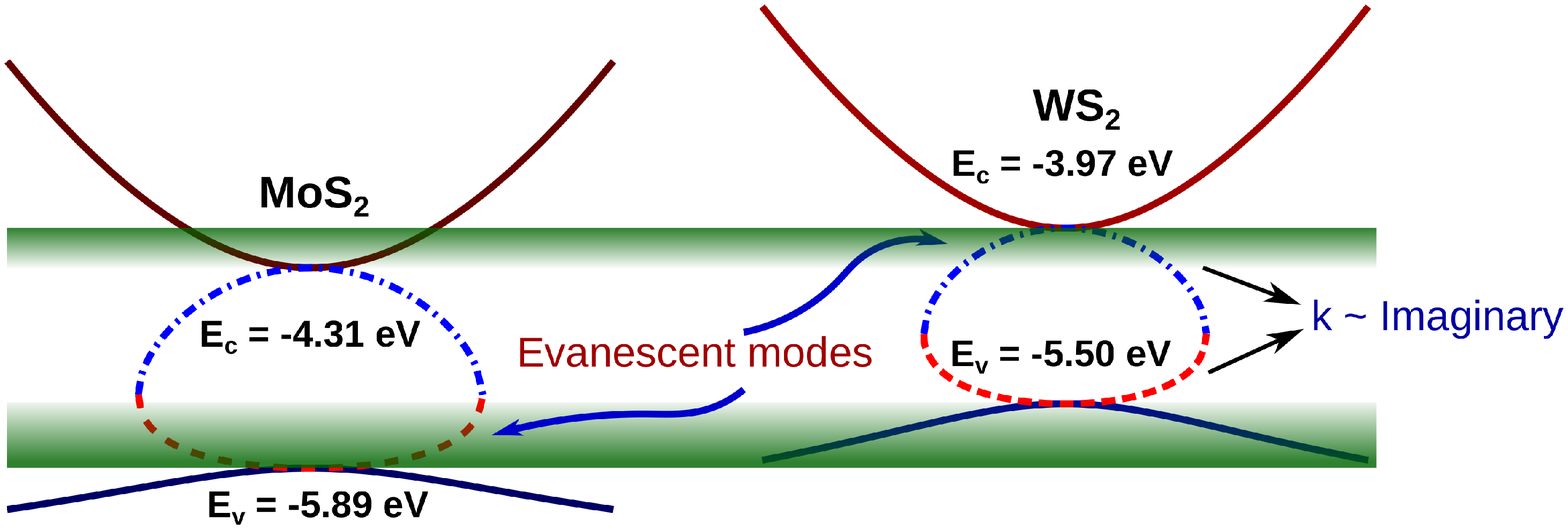}
	\caption{The conduction and valence band near the $K-$point are plotted for MoS$_2$ and WS$_2$ 1L. In the presence of scattering centers, below the energy $-3.97\,$eV (for $n$-type MoS$_2$), and above the energy $-5.89\,$eV (for $p$-type WS$_2$) carrier transport across the interface occurs only through the evanescent bands.}
	\label{fig:bandstructure}
\end{figure}
\begin{figure}[h]
	\centering
	\includegraphics[width = 2.5in]{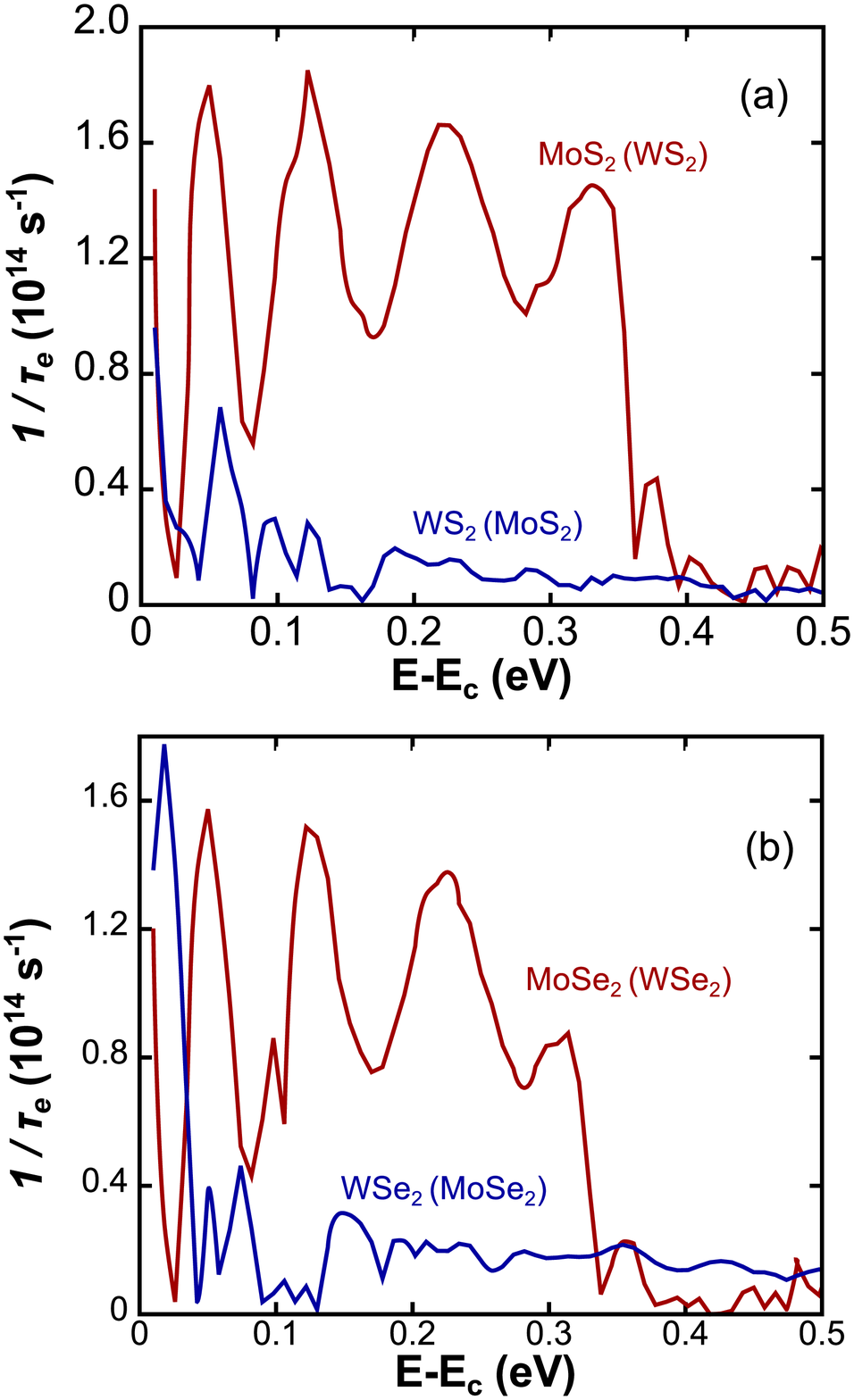}
	\caption{The electron scattering rates versus energy for a $K$-valley electron are plotted for transport in (a) MoS$_2$(WS$_2$) \& WS$_2$(MoS$_2$), and (b) MoSe$_2$(WSe$_2$) \& WSe$_2$(MoSe$_2$) 1L heterostructures. For simplicity, we denote WS$_2$ monolayer with MoS$_2$ inclusions as WS$_2$(MoS$_2$). Same notation applies for all other combinations as well.}
	\label{fig:electronrate}
\end{figure}

To calculate the carrier scattering time $\tau_e$, we have employed a multiscale quantum transport framework informed by first-principles calculations described in Ref.~\cite{sathwikopendomain}. Material inclusions break the translation symmetry of the system. Hence, scattering in these heterostructures can occur via both propagating (real wavevector) and evanescent modes (purely imaginary wavevector). As an example, in Fig.~\ref{fig:bandstructure}, we have plotted the heterointerface formed between 1L MoS$_2$ and WS$_2$. Electronic structure calculations dictate that 1L MoS$_2$ \mbox{($E_c =-4.31\,$eV)} has a lower conduction band (CB) minimum than 1L WS$_2$ \mbox{($E_c = -3.97\,$eV)}. When an electron in the CB with energy \mbox{$-4.31\,$eV$\leq\!\!E\!<\!\!-3.97\,$eV} is injected from 1L MoS$_2$ to WS$_2$, scattering occurs only through the evanescent modes. Similarly, for carriers in the valence band of $p$-type WS$_2$, with energy \mbox{$-5.50\,$eV$\geq \!\!E>\!-5.89\,$eV} (corresponding to the valence band maximum of MoS$_2$) only evanescent modes are available for scattering. Evanescent modes are situated within the bandgap, and result in exponentially decaying contributions to the scattered wavefunction. 

Thermoelectric quantities in Eq.~(\ref{eq:thermoelectric}) are directly proportional to the total scattering time $\tau(E)$. Figure~\ref{fig:electronrate} displays the electron scattering rate, $1/\tau_e$, as a function of energy for four different $n$-type material combinations. We see that WS$_2$(MoS$_2$) has a higher scattering time (inverse of the scattering rate) while compared to MoS$_2$(WS$_2$). A similar trend is followed by WSe$_2$(MoSe$_2$) and MoSe$_2$(WSe$_2$) 1L as well. 

\subsection{Power factor and $ZT$ values in TMDC heterostructures}
The main results for the peak power factor and $ZT$ values for the $n$-type and $p$-type TMDC lateral heterostructures are listed in Table~\ref{tab:ntypezt} and Table~\ref{tab:ptypezt}, respectively. In these tables, the notation A(B) represents that the material B inclusions are confined within the matrix of the material A. The material inclusion is considered here to be an equilateral triangle. 
\begin{center}
\begin{table*}[]
	  \caption{The peak power factor (PF) and the figure-of-merit $ZT$ are listed for $n$-type monolayer (1L) TMDC heterostructures at temperatures $300\,$K, $500\,$K, and $800\,$K. Here, the notation A(B) represents that the material B inclusions are confined within the matrix of the material A. The material inclusion is equilateral triangle of the side length $8\,$nm. The density of inclusions is consider to be, $n_d = 10^{12}\,$cm$^{-2}$. For comparison, we have listed the room temperature $ZT$ values for pristine 1L TMDCs obtained via our scattering calculations along with values reported in the literature Refs.~\cite{Ozbal, HuangTMD1}. }
    \centering
    \renewcommand{\arraystretch}{2.0}
    \begin{tabular}{c | p{1cm} p{1cm} p{1cm} | p{1cm} p{1cm} p{1cm}| c| c| c | c}    
        \hline\hline
       1L heterostructure & \multicolumn{3}{c}{PF ($10^{-3}\,$WK$^{-2}$m$^{-1}$)} & \multicolumn{3}{|c}{$ZT$}&\multicolumn{4}{|c}{Pristine 1L (at 300 K)}\\
        \hline
       A\,(B) &  $300\,$K & $500\,$K & $800\,$K & $300\,$K & $500\,$K & $800\,$K & A&$ZT_{\rm 1L}$& $ZT_{\rm 1L}$ \cite{Ozbal}& $ZT_{\rm 1L}$ \cite{HuangTMD1}\\
       \hline
      MoS$_2$\,(WS$_2$) & 0.365 &0.293 & 0.181 &0.093 & 0.124 & 0.125&MoS$_2$&0.25&0.22 & 0.25\\
      MoS$_2$\,(MoSe$_2$) & 0.335&0.256 &0.167 & 0.084& 0.109 & 0.115&MoS$_2$&0.25&0.22 & 0.25\\
      WS$_2$\,(MoS$_2$) & 4.565 & 3.896 & 2.371 &0.598 & 1.231 & 1.641&WS$_2$&0.20&0.22 & 0.23\\
      WS$_2$\,(WSe$_2$) & 5.977 &4.476 &2.470 & 0.997 &1.611 & 1.806&WS$_2$&0.20&0.22 & 0.23\\
     MoSe$_2$\,(WSe$_2$) & 0.500&0.367 &0.200 &0.173 &0.227 & 0.205&MoSe$_2$&0.38&0.35 & 0.36\\
      MoSe$_2$\,(MoS$_2$) &0.485 &0.362 &0.199 &0.165 & 0.223 &0.205&MoSe$_2$&0.38&0.35 & 0.36\\
     WSe$_2$\,(MoSe$_2$) &1.929 &1.457 &0.815 &0.485 &0.816 &0.875&WSe$_2$&0.30&0.33 & 0.38\\
      WSe$_2$\,(WS$_2$) & 1.954&1.468 & 0.819& 0.488 &0.821&0.879 &WSe$_2$&0.30&0.33 & 0.38\\
      \hline\hline
    \end{tabular}
    \label{tab:ntypezt}
\end{table*}
\end{center}

We observe that the $n$-type WS$_2$(WSe$_2$), and $p$-type MoSe$_2$(WSe$_2$) have the maximum $ZT$ values at room temperature. On the other hand, $n$-type WS$_2$(WSe$_2$), $n$-type WS$_2$(MoS$_2$), and $p$-type MoS$_2$(MoSe$_2$) have larger $ZT$ values at higher temperatures. In Tables~\ref{tab:ntypezt} and \ref{tab:ptypezt}, for comparison, we have listed the room temperature $ZT$ values for pristine 1L TMDCs obtained from Refs.~\cite{Ozbal, HuangTMD1}. For the $n$-type 1L WS$_2$ we observe up to five times larger $ZT$ value with WSe$_2$ inclusions as compared to pristine $n$-type WS$_2$ 1L. Similarly, for $p$-type MoSe$_2$ with WSe$_2$ inclusions, we observe an enhancement by a factor of two in the $ZT$ values as compared to pristine 1L MoSe$_2$. In general, $ZT$ values increase with temperature, as there is a multiplicative factor of temperature in Eq.~(\ref{eq:zt}). 
\begin{center}
\begin{table*}[]
	  \caption{The peak power factor (PF) and the figure-of-merit $ZT$ are listed for $p$-type monolayer (1L) TMDC heterostructures at temperatures $300\,$K, $500\,$K, and $800\,$K. Here, the notation A(B) represents that the material B inclusions are confined within the matrix of the material A. The material inclusion is considered to be an equilateral triangle of the side length $8\,$nm. The density of inclusions is consider to be, $n_d = 10^{12}\,$cm$^{-2}$. For comparison, we have listed the room temperature $ZT$ values for pristine 1L TMDCs obtained via our scattering calculations along with values reported in the literature Refs.~\cite{Ozbal, HuangTMD1}.}
    \centering
    \renewcommand{\arraystretch}{2.0}
    \begin{tabular}{c | p{1cm} p{1cm} p{1cm} | p{1cm} p{1cm} p{1cm}| c | c| c |c}
        \hline\hline
       1L heterostructure & \multicolumn{3}{c}{PF ($10^{-3}\,$WK$^{-2}$m$^{-1}$)} & \multicolumn{3}{|c}{$ZT$}&\multicolumn{4}{|c}{Pristine 1L (at 300 K)}\\
        \hline
       A\,(B) &  $300\,$K & $500\,$K & $800\,$K & $300\,$K & $500\,$K & $800\,$K & A& $ZT_{\rm 1L}$&$ZT_{\rm 1L}$ \cite{Ozbal} & $ZT_{\rm 1L}$ \cite{HuangTMD1} \\
       \hline
      MoS$_2$\,(WS$_2$) &  1.940&1.862 & 1.315 &0.407 &0.713 &0.871&MoS$_2$ &0.50 &0.47 & 0.53\\
      MoS$_2$\,(MoSe$_2$) &4.076  &3.213 & 2.001& 0.648 & 1.115&1.289&MoS$_2$ &0.50 &0.47 & 0.53\\
       WS$_2$\,(MoS$_2$) &0.895  &0.779 &0.568 &0.274 &0.407 & 0.486& WS$_2$&0.40&0.43 & 0.42\\
      WS$_2$\,(WSe$_2$) &1.274  &1.203 &0.873 &0.370 &0.607 &0.736 &WS$_2$ &0.40&0.43 & 0.42\\
      MoSe$_2$\,(WSe$_2$) &2.272  &1.826 &1.060 &0.714 &1.004 &1.045&MoSe$_2$ &0.42 &0.38 & 0.39\\
      MoSe$_2$\,(MoS$_2$) &2.015  &1.554 &0.945 &0.560 & 0.861 & 0.934& MoSe$_2$&0.42 &0.38 & 0.39\\
      WSe$_2$\,(MoSe$_2$) &0.034  &0.023 &0.014 &0.015 &0.017 &0.016&WSe$_2$ &0.35 &0.34 & 0.35\\
      WSe$_2$\,(WS$_2$) & 0.023 &0.017 &0.011 & 0.011 & 0.013 & 0.013&WSe$_2$  &0.35 &0.34 & 0.35\\
     \hline\hline
    \end{tabular}
    \label{tab:ptypezt}
\end{table*}
\end{center}

The calculated peak value of the PF for $n$-type 1L WS$_2$(WSe$_2$) and WS$_2$(MS$_2$) at room temperature is  $5.977\,$mWK$^{-2}$m$^{-1}$ and $4.565\,$mWK$^{-2}$m$^{-1}$, respectively. These values are about twice the peak PF value observed in the corresponding pristine 1L TMDCs \cite{Hippalgaonkar}. Moreover, they are of the same order of magnitude as the observed PF in traditional thermoelectric materials, such as Bi$_2$Ti$_3$ ($5.2\,$mWK$^{-2}$m$^{-1}$ \cite{Satterwaithe}) and BiSbTe ($5.4\,$mWK$^{-2}$m$^{-1}$ \cite{Caillat}) crystals.  

In Table~\ref{tab:ntypezt}, we observe that $n$-type MoS$_2$(WS$_2$) and MoS$_2$(MoSe$_2$) have significantly lower thermoelectric values compared to a pristine MoS$_2$ 1L. Similarly,  $p$-type WSe$_2$(WS$_2$) and WSe$_2$(MoSe$_2$) have significantly lower thermoelectric values compared to a pristine WSe$_2$ 1L (see Table~\ref{tab:ptypezt}). These phenomena can be explained as a direct consequence of band alignment as explained in the next section. 

\subsection{Band alignments and the thermoelectric enhancement}\label{subsec:bandalignment}
\begin{figure}[h]
    \centering
    \includegraphics[width=2.5in]{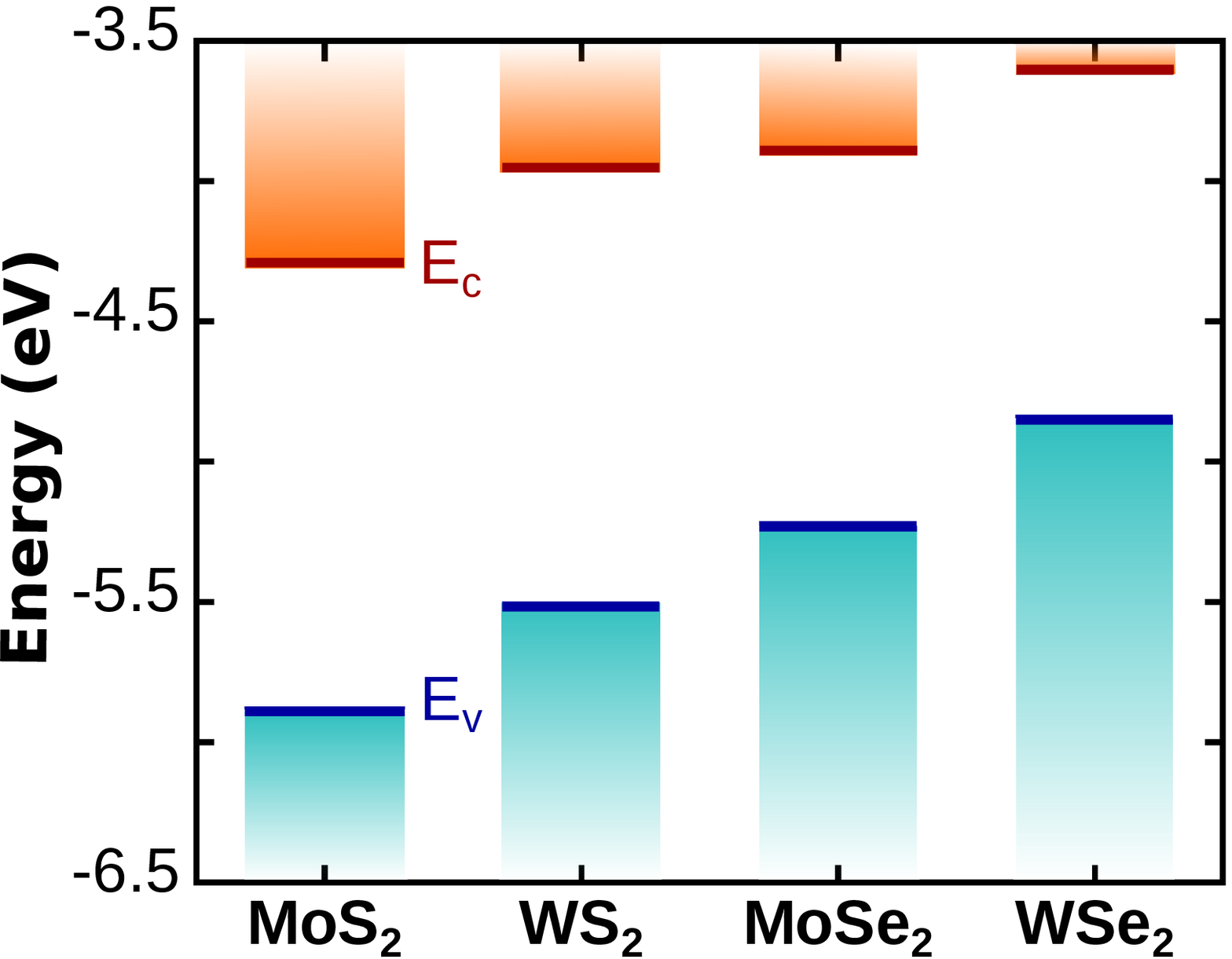}
    \caption{The bar chart displays the band alignment in the semiconducting TMDC 1L. The conduction band minimum and the valence band maximum are represented by $E_c$ and $E_v$, respectively. These numerical values are listed in Table~\ref{tab:parameters}.}
    \label{fig:bandalignment}
\end{figure}
In Fig.~\ref{fig:bandalignment}, we observe that 1L WS$_2$ has a higher CB minimum at the $K$-valley than 1L MoS$_2$. Hence, MoS$_2$ inclusions provide additional conduction channels for electrons entering from the $n$-type 1L WS$_2$. This will increase the electron scattering time as seen in Fig.~\ref{fig:electronrate}. On the other hand, in the $n$-type MoS$_2$ with WS$_2$ inclusions, scattering occurs through the evanescent modes offered by the WS$_2$ inclusion. This will significantly decrease the scattering time. Moreover, being real functions the evanescent modes indirectly decrease the probability current by draining the probability of propagating channels, thereby, significantly reducing the conductance values. 

The total scattering time $\tau(E)$ follows the reciprocal sum rule defined in Eq.~(\ref{eq:Matthiessen}). Hence, the lower of the two scattering times between $\tau_{ph}$ and $\tau_{e}$ will be the dominating contributor to the thermoelectric quantities. For transport in $n$-type MoS$_2$(WS$_2$) 1L, $\tau_{e}$ is an order of magnitude lower than $\tau_{ph}$ around the band edge. Hence, we obtain low values of PF and $ZT$ values as shown in Table~\ref{tab:ntypezt}. 

\begin{figure*}[t!]
    \centering
    \includegraphics[width = 5in]{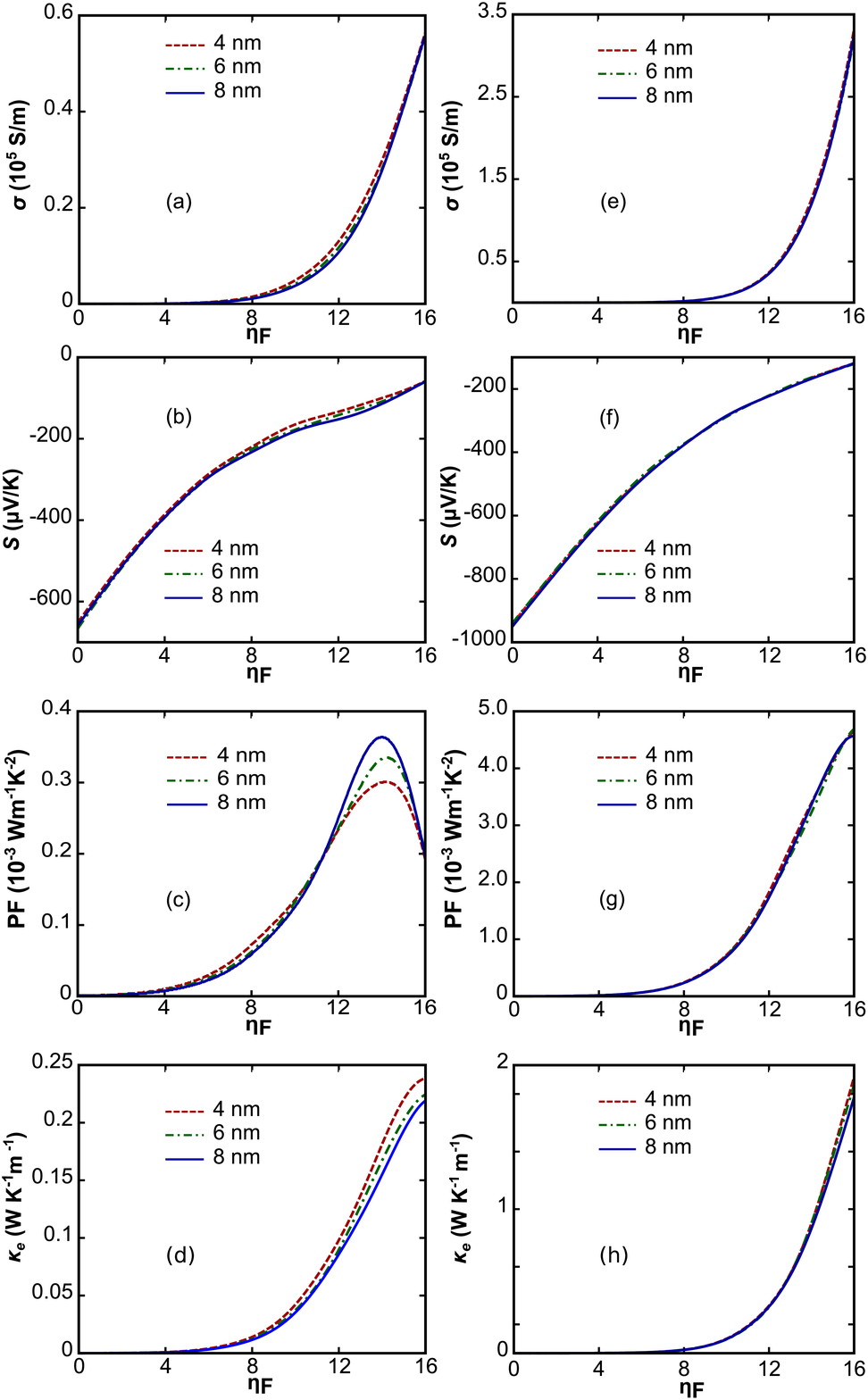}
    \caption{The conductance, Seebeck coefficient, power factor, and electrical-thermal conductivity are plotted as a function of the reduced Fermi-level $\eta_F = (E-E_c)/k_B T$ for the  $n$-type MoS$_2$(WS$_2$), and $n$-type WS$_2$(MoS$_2$) 1L heterostructures. We have considered triangular material inclusions of radius $4, 6,$ and $8\,$nm.}
    \label{fig:thermoelectricM2WS}
\end{figure*}
In $p$-type heterostructures, the valence band (VB) maxima between the two layers determine the occurance of evanescent modes. In Fig.~\ref{fig:bandalignment}, we see that WS$_2$ has a higher VB maximum than MoS$_2$. Thus scattering in the $p$-type WS$_2$(MoS$_2$) occurs through evanescent modes. Hence, we observe that the $p$-type WS$_2$ heterostructures have lower $ZT$ values while compared to the corresponding pristine monolayer as seen in Table~\ref{tab:ptypezt}. Due to additional conduction channels offered by the WS$_2$ inclusions, an enhancement in the $ZT$ and PF values are observed in the $p$-type 1L MoS$_2$(WS$_2$). An analogous mechanism explains the enhancement observed in $p$-type MoSe$_2$(WSe$_2$), which has the highest room temperature $ZT$ value amongst the $p$-type heterostructures.     

Figure~\ref{fig:thermoelectricM2WS} displays the conductance, Seebeck coefficient, PF, and electrical thermal conductivity as a function of the reduced Fermi-level $\eta_F = (E-E_c)/k_B T$ for $n$-type MoS$_2$(WS$_2$), and $n$-type WS$_2$(MoS$_2$) 1L heterostructures. Note that $\sigma$ and $\kappa_e$ monotonically increase with $\eta_F$. Comparing Fig.~\ref{fig:thermoelectricM2WS}(a) and (e), we observe a significant enhancement in $\sigma$ for WS$_2$(MoS$_2$) as reasoned earlier, whereas the Seebeck coefficient remains of the same order for both the heterostructures. Hence, we obtain high PF and $ZT$ values for the $n$-type 1L WS$_2$(MoS$_2$) heterostructures. A similar trend is followed by other material combinations as well. We note that the PF increases slightly with increase in the width of material inclusions, and we found it to be optimized for the side length of $8\,$nm for the equilateral triangle inclusions. Unlike Schr\"odinger particles, the massive Dirac particles in TMDC inclusions have a critical length below which they will not occupy any bound states \cite{shenoy}. Thus, for inclusions with edges smaller than this critical length, heterostructures considered here will have the thermoelectric values similar to that of  \mbox{1L pristine.}    

\section{Conclusions}\label{sec:conclusions}

In this work, we investigated the thermoelectric properties of semiconducting transition metal dichalcogenide lateral heterostructures using a multiscale quantum transport framework. We reported a new mechanism to enhance the thermoelectric efficiency in 2D materials by adding conduction channels through lateral heterostructures.  
The $n$-type WS$_2$ monolayer with WSe$_2$ inclusions has the highest room-temperature $ZT$ values, which is about five times larger than the pristine WS$_2$ monolayer. The $p$-type MoSe$_2$ monolayer with WSe$_2$ inclusions is observed to have a room-temperature $ZT$ value about two times larger than the pristine MoSe$_2$ monolayer. The peak PF values calculated in these heterostructures are of the same order of magnitude as traditional high-performance thermoelectric materials such as Bi$_2$Ti$_3$ and BiSbTe. The PF values reported here, are also the highest amongst gapped 2D monolayers. 
We expect the given mechanism to show similar enhancement of thermoelectric power in oxide and other monolayer heterostructures. Hence, 2D lateral heterostructures provide exciting new avenues to develop ultra-efficient in-plane thermoelectric devices.     
\section{Methods}\label{sec:calculations}
To determine the thermoelectric quantities in Eq.~(\ref{eq:thermoelectric}) we need to determine the density of states $g(E)$ and the total scattering time $\tau(E)$. The density of states $g(E)$ is computed from DFT calculations. The phonon and electron scattering contributions of $\tau(E)$ are determined independently.

DFT calculations were performed, using the Vienna Ab Initio Simulation Package (VASP) \cite{kresse1, kresse2}, to obtain the density of states for the various TMDC monolayers studied here. Core and valence electrons were modeled using the projector-augmented wave (PAW) method \cite{blochl, kresse3} and the local density approximation \cite{ceperley, perdew} was used to describe electron exchange and correlation. The kinetic energy cutoff was set to $500\,$eV and a Gaussian smearing of $0.1\,$eV was used for Brillouin-zone integrations. The Brillouin zones were sampled using an \mbox{$18\times18\times1$} \mbox{$\Gamma$-centered} \mbox{$k$-point} mesh. The lattice parameters for the TMDC monolayers were fixed at the bulk experimental parameters (MoS$_2$: \cite{moss}; MoSe$_2$: \cite{lavik}; WS$_2$ \& WSe$_2$: \cite{schutte}).  As semilocal functionals underestimate the fundamental gaps of semiconductors, the density of states were shifted, {\it a posteriori}, to match the quasiparticle gaps reported in Ref.~\cite{C2DB}. 

The deformation potentials obtained from first-principles calculations \cite{Kaasbjerg, kim_eph} were employed to compute both the acoustic and optical phonon mode contributions. We have included the zeroth order acoustic and optical, and first order acoustic mode contributions. The effect of the Fr\"ohlich interaction is implicitly added to the deformation potential \cite{kim_eph}. The total $\tau_{ph}$ is determined using the reciprocal sum of these contributions as prescribed by Matthiessen’s law. Further details of these calculations can be found in Ref.~\cite{sathwikopendomain}. 

For elastic scattering processes in 2D, the kinetic definition of $\tau_e$ is given by $\tau_{e}(E) = {1}/({n_d \sigma_m \left<v\right>}),$
where $n_d$ is the disorder density, $\left<v\right>$ is the average velocity, and $\sigma_m$ is the momentum scattering cross-section. $\left<v\right> = 2v/\pi$ for a uniform incoming velocity distribution. $\sigma_m$ is determined using a multiscale quantum transport framework \cite{sathwikopendomain}. Carrier scattering in lateral heterostructures occurs via both propagating and evanescent modes. Standard scattering calculations cannot account for these crucial contributions of decaying evanescent modes, since their probabilities vanish at the asymptotic limit, where the boundary conditions are applied to determine the scattering amplitudes. On the other hand, our methodology accurately includes these contributions since we have circumvented the need for the asymptotic boundary conditions by introducing absorbers around the scattering centers.

\begin{table*}[t]
    \centering
   \begin{tabular}{c c c c c c c c c c c}
      \hline\hline & & & & & & & & & &\\
      & & $E_c$  & $E_v$ & $a$  & $t$ & $\lambda$ & $\alpha_{+}(\alpha_{-})$ &$m^c_K$ & $m^v_K$& \\[5pt]
      & & (eV)& (eV) &($\angstrom$) &(eV) &(eV) &eV$\cdot$\angstrom$^2$ & ($m_e$)&($m_e$) &\\
      \hline
      & & & & & & & & & &\\
      & MoS$_2$ &-4.31 &-5.89 &3.184 &1.059 &0.073 &-5.97(-6.43)&0.45 &0.53 & \\[5pt]
      & WS$_2$ &-5.97 &-5.50 &3.186 &1.075 &0.211 &-6.14(-7.95) &0.3 &0.34 & \\[5pt]
      & MoSe$_2$ &-3.91 &-5.23 &3.283 &0.940 &0.090 &-5.34(-5.71) &0.53 &0.58 &\\[5pt]
      & WSe$_2$ &-3.61 &-4.85 &3.297 &1.190 &0.230 &-5.25(-6.93)&0.33 &0.36 & \\[5pt]
      \hline\hline
   \end{tabular}\label{tab:parameters}
    \caption{The band parameters used in our calculations are listed. These parameters were obtained from the previously reported first-principles study \cite{C2DB, kormanyos1}.}
\end{table*}
Scattering properties are determined through a $2$-band $\textit{\textbf{k}}\cdot\textit{\textbf{p}}$ Hamiltonian of the form
\begin{equation}\label{eq:hkptmd}
H_{kp} = H_0 + at\left(\eta\, k_x \hat{\sigma}_x + i k_y \hat{\sigma}_y\right) - \lambda\,\eta \frac{(\hat{\sigma}_z - 1)}{2} s,
\end{equation}
where $\hat{\sigma}$ denotes the Pauli matrices, $a$ is the lattice constant, $t$ is the effective hopping integral, $\lambda$ is the spin-orbit (SO) parameter, and $H_0$ is given by
\begin{equation}
H_0 = \displaystyle\left[\begin{array}{cc}
   E_c + \alpha_{s} k^2  & 0 \\
   0   &  E_v 
\end{array}\right].
\end{equation}
Here, $\alpha_{s}$ is a material parameter, and $E_c$ and $E_v$ are the CB minimum and the VB maximum at the $K$-valley, respectively. All the material parameters employed in this article are listed in Table~\ref{tab:parameters}. A detailed discussion of the quantum transport framework employed here is described in Ref.~\cite{sathwikopendomain}. The quantum transport calculations combined with the Boltzmann transport theory determine $\sigma$, $S$ and $\kappa_e$ (see Eq.~(\ref{eq:thermoelectric})). To determine the $ZT$ factor we also require the phonon thermal conductivity $\kappa_{ph}$ (see Eq.~(\ref{eq:zt})). Typically $\kappa_{ph}$ is computed through the phonon dispersion relations. We have utilized the $\kappa_{ph}$ values listed in Ref.~\cite{Ozbal}, that are obtained using DFT calculations.
\begin{acknowledgments}
Calculations presented here were performed using computational resources supported by the Academic and Research Computing Group at WPI.
\end{acknowledgments}

\end{document}